# TCLK MUST STAY! CAMAC MUST GO! HOW DOES FERMILAB MOVE FORWARD*

M. R. Austin[†,1], L. Carmichael[1], D. McArthur[1], E. Milton[1], and A. Quilty[1]
[1]Fermi National Accelerator Laboratory, Batavia, USA

*Abstract*

The current Timing System at Fermilab has been around for 40 years and currently relies on 7 CAMAC crates and over 100 CAMAC cards to produce the Tevatron Clock (TCLK). Thanks to the ingenuity of those before us, this has allowed Fermilab the flexibility to change the timing and Events for its accelerator as beamlines and projects have changed over the years. With the advent of the Proton Improvement Plan-II (PIP-II), the Timing System at Fermilab is being reimagined into a single chassis with even greater flexibility and functionality for decades to come while tackling the ever-challenging task of maintaining backwards compatibility.

## INTRODUCTION

TCLK at Fermilab is used to transmit important accelerator timing information to all major systems throughout the accelerator complex with up to 256 unique triggered 8-bit "Events" that are used as timing markers. TCLK is a 10 MHz serial signal which uses an encoding scheme called bi-phase or modified Manchester coding to transmit 10 bits (1 start bit, 8 data bits, and 1 parity bit). There are then two additional 1's transmitted before the next Event is allowed to be transmitted [1]. TCLK begins with a 10 MHz source and the Master Clock Oscillator (MCO). The MCO triggers synchronous to the 60 Hz line frequency including the trigger for the Timeline Generator (TLG) to play out certain Events that are generally related to Machine Resets. Some of these Machine Resets then trigger other hardware that triggers the playing out of other Events. The timing of all these Events is based on programmable delays, external Event requests from remote hardware, as well as delays that synchronize Event requests to the 60 Hz line frequency. To accomplish all of this, there are 7 Computer Automated Measurement and Control (CAMAC) crates and over 100 cards with settable devices for about half the cards and hard coded settings for the other half, a few custom chassis, and a VME Front End for the TLG interface. The hardware involved with creating TCLK is shown in the collage of CAMAC Crates in Fig. 1.

## MOTIVATION

PIP-II necessitates the first major modification to Fermilab's timing system since the Tevatron in the 1980s. The new system must provide inter-bunch precision and provide full backwards compatibility with the rest of the complex. This means that the backwards compatible TCLK output from the Accelerator Clock (ACLK) Generator that will interface with older hardware, needs to look and act exactly like the existing TCLK signal. The new system must meet numerous specifications for user interface, Permit inputs, Event triggers, and enabling, disabling, and Timestamping of Events. This results in two systems being necessary: The ACLK Generator, to provide general machine timing and the Linac Clock (LCLK) Generator to provide beam synchronous timing to PIP-II.

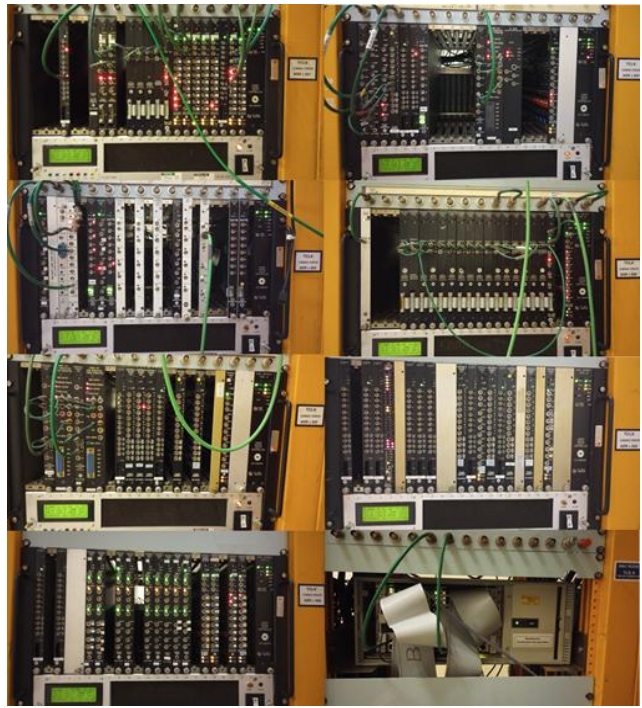

Figure 1: CAMAC hardware for creating TCLK.

## SPECIFICATIONS

Given the similarity between the needs of ACLK and LCLK, it was determined that a system could be designed to share common physical hardware and logic implementation. This system is colloquially known as a Clocking System. This common hardware is shown in Fig. 2. A significant portion of the logic on the inside is similar as well. The Clocking System consists of:

- A System-on-Module (SOM) with an Arria 10 Field Programmable Gate Array (FPGA) containing a Hard Processor System (HPS) to handle all network communication for the user interface. This includes setting devices for various Event triggers and delays, loading new Timelines and Machine Data (MDAT) settings [2].

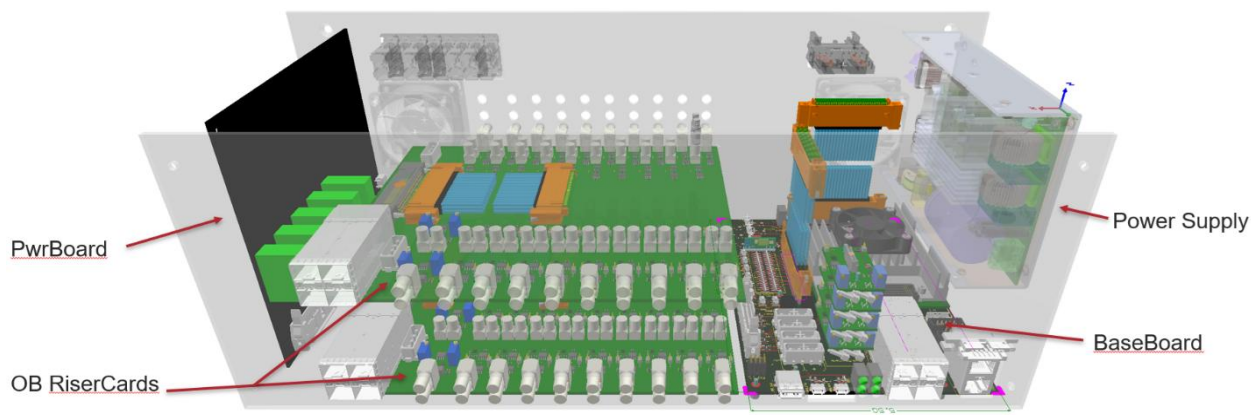

Figure 2: Clocking System – ACLK Generator or LCLK Generator

- The ACLK and LCLK signals are transmitted utilizing transceivers within the FPGA on Single Mode (SM) Fiber using 8B10B Encoding with continuous data transmission. Each frame of data contains 16-bit Event information and 64-bits of Data for a timestamp, status bits, or some other pertinent information.
- The ACLK data rate is 1.2 Gbps using a GPS based 10 MHz source while the LCLK data rate is a beam synchronous 1.21875 Gbps using a 10.1562 MHz clock based on the PIP-II Low-Level Radio Frequency (LLRF).
- 80 I/O channels, each capable of up to 200 MHz, are available on the rear panel to be used for input permits, input triggers for asynchronous Event generation, status and trigger outputs, and future undefined I/O. There are also duplicate status outputs on the front of the Clocking System for monitoring all the rear I/O.
- The ACLK Generator also provides a backwards compatible version of TCLK for use until further upgrades at Fermilab take place.
- The 10 MHz edge of the start-bit of a TCLK Event will simultaneously begin the transmission of the corresponding ACLK Event and Data.
- Repetition rate for Booster Beam Resets is upgraded to 20 Hz instead of 15 Hz.
- Extensive Interface Specifications were created to guide the design so that the final system would meet all the requirements of the stakeholders and allow for future development.

## CLOCK FRAME

ACLK and LCLK have frames consisting of a 96-bit payload as shown in Fig. 3. The use of 8B10B encoding results in a 120-bit serial data stream.

| 16-Bit EVENT | 64-Bit Data | 8-Bit CRC | Comma | (96-Bits)

Figure 3: ACLK and LCLK frame definition.

With ACLK operating at 1.2 Gbps, the 120-bit serial data stream encodes an Event every 100 nS in line with the 10 MHz clock. This works out quite nicely with 10 MHz rate of TCLK, allowing for backwards compatibility. The existing TCLK meanwhile, has a minimum Event spacing of at least 1.2 uS. Therefore, there are at least 12 ACLK data frames between any TCLK Event sent. This allows for the ACLK Generator to retain synchronicity between the two encoded clocks.

Meanwhile, with LCLK operating at 1.21875 Gbps, the 120-bit serial data stream encodes an Event about every 98.46 nS in line with the 10.1562 MHz clock. This is a factor of 16 in relation to the 162.5 MHz RF used for the PIP-II Super Conducting Linac [3].

## TIMELINE GENERATOR

The TLG consists of a VME board with hardware and software that the operator interacts with for building a Timeline for the operation of the Accelerator [2]. Moving forward, the hardware aspect of the TLG that plays out the requests for Events to be encoded will reside within the Clocking System. The software that the operator interacts with that calculates and builds the Timeline will exist in a web-based Flutter application and a GraphQL API. This application then interacts with the Input Output Controller (IOC) running on the HPS of the SOM to read and write the settings for the Timeline as well as all the other device settings.

The Timeline is set up to have up to 6000 programmable "Slots". Each Slot has 8 groups of 8 different 8-bit Events. These groups are for the various Events that serve as Machine Resets. The Machine Resets are used for preparing that particular part of the accelerator for beam for a given slot. The Timeline also contains 16 different 8-bit Machine Data (MDAT) Type Codes and their respective 16-bit MDAT Data. The MDAT Type Codes and Data transmit information about the current status of various aspects of the machine. All this information makes up the Timeline. The operator has the ability to adjust the length of the Timeline by selecting the number of Slots desired. Once the HPS loads the Timeline into the on-chip Random Access Memory (RAM) within the FPGA, the FPGA takes over and goes through the sequence of reading each slot in memory at a 20 Hz rate. The 6000 Slots at 20 Hz results in a 5-minute maximum length for a Timeline. There are a total of 4 groups of RAM within the FPGA that hold Timeline information. Two different "Scheduled" Timelines and two different "One-Shot" Timelines. The operator has the option to select the existing Scheduled Timeline, load the next Scheduled Timeline, perform a One-Shot with the currently selected One-Shot Timeline, or perform a One-Shot with the next One-Shot Timeline. A Scheduled Timeline will continue to repeat the set number of Slots repeatedly encoding TCLK and ACLK. While a One-Shot Timeline will encode TCLK and ACLK only going through the set number of Slots once. Upon completion of a One-Shot Timeline, the Timeline reverts to the original Scheduled Timeline where it is then repeated continuously.

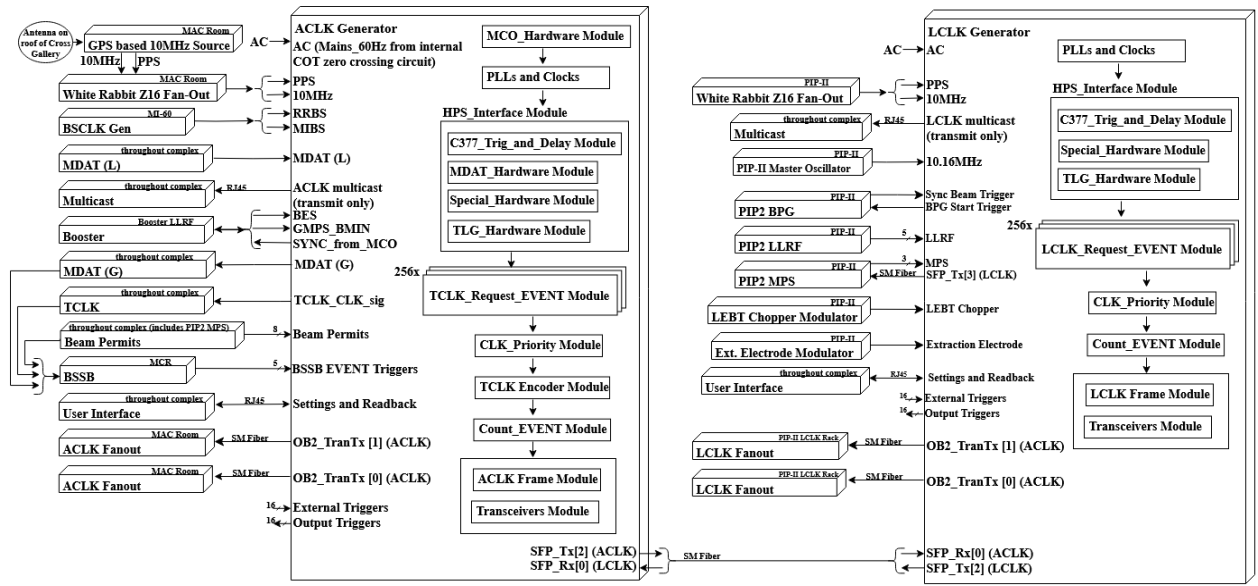

Figure 4: ACLK and LCLK Generator System Diagram.

Once the Timeline is all loaded, the system will continue to operate without any operator intervention. When the Timeline reaches the end or the last Slot, the first Event of the Timeline is encoded again, which is known as the Super Cycle Reset. This signifies the beginning of the Timeline.

The Timeline Events are set to play out with multiple programmable delays initiated from the 20 Hz frequency that was derived from the 60 Hz electric utility provider. This must remain the case due to the nature of the way the Booster accelerator operates.

The functionality of the TLG and its Timeline is a part of what has given Fermilab the flexibility to change the final destination of the beam and update various accelerators paving the way for multiple physics discoveries.

## TRIGGERED EVENTS

The Timeline created by the TLG only defines about 30% of the Events that are currently encoded onto the clock. The other Events are set up as "Triggered" Events. The encoding of these can be triggered from external inputs from other hardware, input permits, "Reflected" Events from another encoded clock, or the operator can set up Events to watch the clock to see when a different Event just played out to then encode the desired Event. All of these Triggered Events can have programmable delays associated with them. There are also special Events that take extra logic to implement due to the nature of the creation of that Event.

Given the complexity of the logic and interconnects of the various CAMAC hardware that the Clocking System was being utilized for in making the ACLK Generator work, the logic design was made as flexible as possible for future changes.

Although the new ACLK and LCLK signals have 16-bit Events, in order to maintain the backwards compatibility with TCLK, the system is only designed to use the lower 8-bits as the Triggered Events. The upper 8-bits of the 16-bit Events are reserved for informational data and Reflected Events from other clocking systems. This includes when ACLK and LCLK Events are being reflected or encoded onto each other's data stream.

With each Clocking System able to handle up to 256 different Triggered Events, the operators have the ability to enable/disable Events, set delays with respect to when the Event is encoded, and redefine what triggers an Event to become encoded. This flexibility meets the needs and desires of the stakeholders in operating the accelerator complex. The overall system design for the ACLK Generator with its external I/O and various modules to implement this design is shown in Fig. 4.

## TIME TO SYNC

The existing TCLK signal is generated from a standalone 10 MHz source. The 10 MHz source moving forward is now a GPS based system with a 10 MHz and Pulse Per Second (PPS) output being fed into a White Rabbit Z16 Precision Time Protocol (PTP) Timing System. After that, the 10 MHz and PPS are fed directly into the ACLK Generator. These signals are also fed elsewhere around Fermilab through the PTP distribution of the White Rabbit Network. Since the ACLK Generator will now be using the 10 MHz source to generate all its Phase Lock Loops (PLLs), the system can provide Timestamps for certain Events within the 64 bits of Data in the ACLK

Frame. These Timestamps will have the option of being stamped either when the trigger comes into the system or when the Event is encoded. It also means that the systems that have not been updated yet and still use the older TCLK signal, that does not contain any Timestamps within the Data transmitted, will still be able to reference back to data to calculate what the Timestamp was. The Timestamps with the 64-bit data field will be formatted to use the upper 32 bits for the Epoch time in seconds updated from the PPS counter from the White Rabbit System. The lower 32 bits will be a counter that is reset on every PPS that counts using a PLL clock of 100 MHz that is synchronous to 10 MHz signal fed from the White Rabbit Network. This allows for a granularity of 10 nS for the Timestamp. This will meet the needs of various stakeholders for timing and data collection.

## DECODING

For those wishing to have hardware that takes in either ACLK or LCLK and does not rely on the older TCLK, an existing timing system referred to as the Multi-Function Timing Unit (MFTU) has been upgraded to handle ACLK or LCLK. This system is referred to as the MFTU2 and is shown in Fig. 5. This system decodes various programmable Events from the encoded clock and starts counting out programmable delays to then fire output triggers upon completion of the counting of the delay that is synchronous with the clock. The system also has a programmable pulse width for the output with output channels that can be tuned to single nanosecond delays. These types of output triggers are used throughout the Fermilab complex for various ways of controlling the beamline and collecting data.

In order to accommodate users who wish to bring their own hardware for use, a Clock Decoder IP Core is provided by Fermilab to decode the ALCK or LCLK signal. This Core is shown in Fig. 6.

## TESTING

In order to test the backwards compatibility of the ACLK Generator for generating TCLK, the system will be systematically tested for each Event and compared with the existing way TCLK is generated. This method will confirm that all Events are being generated properly and in the correct order. Various Timelines will be set up and tested to confirm that the system plays the Timeline as designed. The user settings will be tested to confirm that Events are able to be enabled/disabled, settings of delays with respect to when the Event is encoded occurs properly, and redefining what triggers an Event to become encoded takes effect correctly.

The upgrade from operating at 15 Hz to 20 Hz will come with its own challenges. This will involve changing the delays for several of the Events that are triggered by other Events that were encoded onto the clock. The operators will need to manually adjust these values to find out the proper settings, so the subsequent Events are encoded at the proper time.

The ACLK and LCLK Generators will also be set up to have multiple MFTU2s set up to decode an Event, count a delay, and fire an output trigger. The outputs of multiple units will be compared for jitter between them. It is necessary for both the ACLK and LCLK Generators to be connected to each other to confirm that the other system's Events are being reflected properly. The ACLK Generator will also be set up to send the proper Events to the LCLK Generator to start the sequence for the LCLK Generator to request beam.

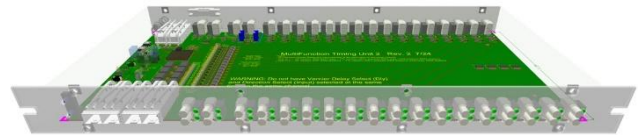

Figure 5: MFTU2 – Used for decoding ACLK or LCLK signals and triggering on Events with a synchronized time delay.

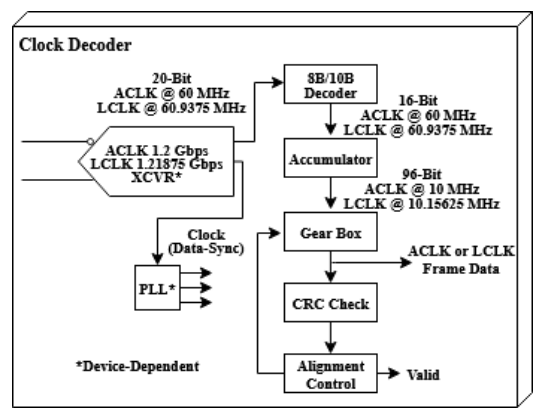

Figure 6: Fermilab IP Core Clock Decoder.

## CONCLUSION

Fermilab's PIP-II project requires extensive upgrades to the Timing System. The existing system simply could not meet the needs for Fermilab to move forward into the future. The Clocking System was designed to create a system that handles the arduous task of backwards compatibility with TCLK, create a timing system for the new PIP-II accelerator, and update the timing system for the rest of Fermilab so it is ready to move forward. It is rigid enough to meet all the demands that PIP-II has placed on the design but flexible enough to meet the unknown needs of the future.